 \documentclass[reprint,secnumarabic,amssymb, nobibnotes, aps, prd, superscriptaddress]{revtex4-1}
\usepackage[T1]{fontenc} 
\usepackage{graphicx}
\usepackage{color}
\usepackage{amsmath}
\usepackage{upgreek}
\usepackage{gensymb}
\usepackage[left,modulo]{lineno}
\usepackage{xcolor}
\usepackage{hyperref}

\begin{document}
\title  {Transition from viscoelastic to fracture-like peeling of pressure-sensitive adhesives}

\author{Marion Grzelka}
\email{m.grzelka@uva.nl}
\affiliation{Van der Waals-Zeeman Institute, Institute of Physics, University of Amsterdam, 1098XH Amsterdam, The Netherlands}
\author{Stefan Kooij}
\affiliation{Van der Waals-Zeeman Institute, Institute of Physics, University of Amsterdam, 1098XH Amsterdam, The Netherlands}
\author{Sander Woutersen}
\affiliation{Van't Hoff Institute for Molecular Science, University of Amsterdam, 1098XH Amsterdam, The Netherlands}
\author{Mokhtar Adda-Bedia}
\affiliation{Universit\'e de Lyon, Ecole Normale Sup\'erieure de Lyon, Universit\'e Claude Bernard, CNRS, Laboratoire de Physique, Lyon F-69342, France}
\author{Daniel Bonn}
\affiliation{Van der Waals-Zeeman Institute, Institute of Physics, University of Amsterdam, 1098XH Amsterdam, The Netherlands}

\begin{abstract}

We investigate the process of the slow unrolling of a roll of a typical pressure-sensitive adhesive, Scotch tape, under its own  weight. Probing peeling velocities down to nm/s resolution, which is three orders of magnitudes lower than earlier measurements, we find that the speed is still non-zero. Moreover, the velocity is correlated to the relative humidity. A humidity increase leads to water uptake, making adhesive weaker and easier to peel. At very low humidity, the adhesive becomes so stiff that it mainly responds elastically, leading to  a peeling process akin to interfacial fracture. We provide a quantitative understanding of the peeling velocity in the two regimes.\\

Published in Soft Matter\\
\doi{10.1039/D1SM01270C}

\end{abstract}
\maketitle

\section{Introduction}
Adhesion is important for many every day, engineering and biological processes, but remains ill-understood from a fundamental level. Different adhesion mechanisms such as mechanical interlocking and electrostatic, chemical and van der Waals bonding have all been proposed~\cite{israelachvili_intermolecular_2015}. However there is no unified theory for adhesion, and many adhesion mechanisms are believed to be specific to particular material combinations. In addition, adhesion forces can depend very sensitively on the specific geometry of the debonding, the peeling force differing by orders of magnitude for the same adhesion energy~\cite{israelachvili_intermolecular_2015}. This makes it notoriously difficult to predict the adhesion behavior. One of the key examples here are pressure sensitive adhesives (PSAs) typically used in adhesive tape and sticky notes~\cite{creton_pressure-sensitive_2003}. In spite of the fact that these are materials that many people use every day, there is no fundamental understanding of the adhesive strength and consequently the force necessary to undo the adhesive bond~\cite{creton_fracture_2016}.
 
In this paper we provide such an understanding for the unsticking of PSAs under different environmental conditions. In these and many other adhesive systems the adhesive is typically 'soft', i.e., visco-elastic~\cite{kaelble_theory_1964,gent_adhesion_1969,derail_relationship_1998,chopin_nonlinear_2018} and this turns out to provide the key to a quantitative understanding. Most of us have experienced sticking the end of a piece of scotch tape to the edge of a table or desk, while using a freshly cut bit from the roll. The generic observation is that the tape stuck to the table does not appear to unroll under the weight of the roll. Contrary to this idea, we show here that at long time scales the tape does in fact start to unroll right away, with a speed scaling with the force, i.e. the weight of the remaining tape on the roll.  We find that the adhesive properties strongly depend on the environment, notably on the humidity, with a very strong dependence of the peeling speed on the environmental humidity. At very low humidity, the PSA becomes very rigid and exhibits a solid-like elastic behavior. In this second regime, we suggest that the unsticking is due to an interfacial fracture propagating with a speed depending on the fracture energy. 

\section{Methods and results}

\begin{figure*}[ht!]
    \centering
    \includegraphics[width=\textwidth]{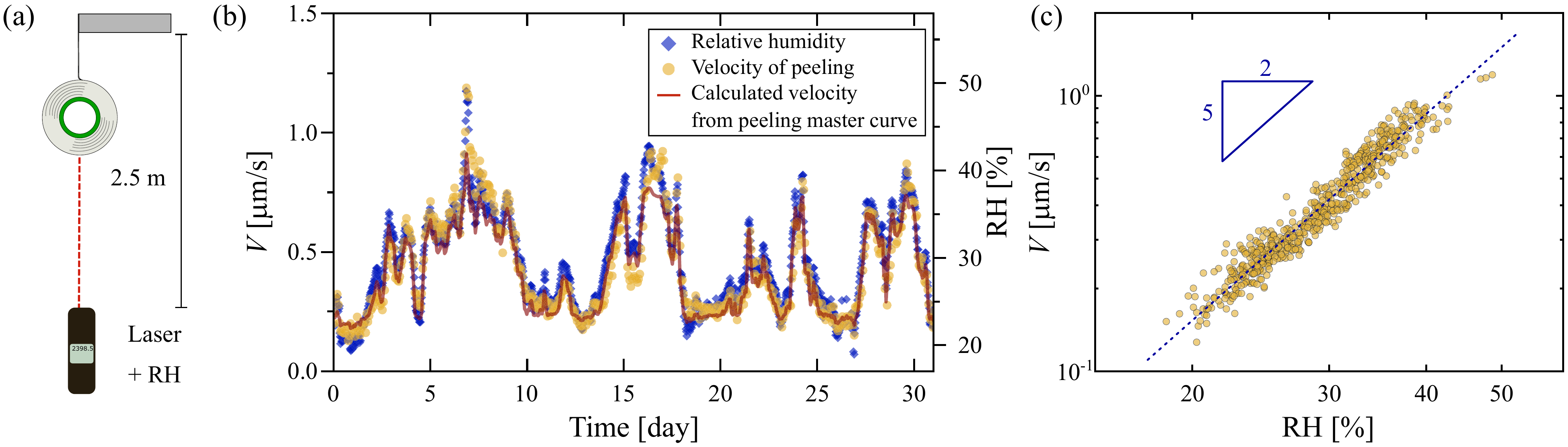}
    \caption{Unrolling of a suspended roll of tape over a period of one month. (a) Illustration of the set-up. A roll of Scotch tape ($m\sim 21\,\mathrm{g}$) is suspended 2.5 m above a reference point, with the exact distance  measured as a function of time using a laser distance meter. Temperature and relative humidity (RH) are monitored throughout the experiments. (b) Both the downward velocity $V$ of the roll of tape and the relative humidity RH strongly fluctuate with time. The red line is the velocity of peeling calculated with the time-humidity superposition principle. (c) Peeling velocity, as extracted from (b), as a function of relative humidity. The dotted line shows a fitted power law of exponent $5/2$.\footnotesize }
    \label{fig:setup_illustration}
\end{figure*} 

The studied tape (3M 810  Magic Scotch) is a PSA composed of a synthetic acrylic adhesive layer of thickness $e= 28\,\upmu$m, and a $38\,\upmu$m thick matte cellulose acetate backing. To investigate the ease with which this tape can be peeled from itself, we first study the unrolling of a suspended roll of tape of approximately $21\,$g under the action of gravity over a period of one month. The changing mass of the roll of tape is less than $1.3\,$g over a month, leading to a change of $6\,\%$ on the applied force: we neglected this change in further discussion. The time of flight of a laser pulse is used to measure the vertical distance from the roll of tape to a reference point (see Fig.\,\ref{fig:setup_illustration}(a)). The setup is surrounded by a metal casing to prevent any airflow from disturbing the tape roll. However, the casing is not isolated from lab fluctuating atmospheric conditions. Throughout the experiments, we monitor the relative humidity RH and temperature (Testo 560).

The peeling velocity $V$ of the tape displays large fluctuations over the measurement period (Fig.\,\ref{fig:setup_illustration}(b)) that are strongly correlated with variations in the relative humidity (RH), with higher RH resulting in faster peeling of the tape. A close inspection of the time-dependencies of $V$ and RH reveals that changes in $V$ are delayed by roughly one hour with respect to changes in the RH, suggesting a time-dependent water uptake by the hygroscopic adhesive layer. Note that the fluctuations of RH over this time scale are negligible ($<1.7\%$). Quantitatively, Figure~\ref{fig:setup_illustration}(c) shows that the velocity scales as a power law with the RH with an exponent of $5/2$: the higher the RH, the lower the resistance of the PSA, leading in turn to faster unrolling of the tape.

Next, we investigate the effect of the  peeling force. In addition to the peeling of the roll of tape, where we attach different weights to the suspended roll, we also tape two layers of the adhesive over each other on a glass plate and peel the upper layer away from the lower layer with a controlled force, mimicking what happens during unrolling of a roll of Scotch tape. The peeling velocity is measured by tracking the peeling front with a CCD camera (Nikon D850 equipped with a
macro-lens Laowa 25 mm 1:2.8 and Phantom Miro M310 high-speed camera with a
macro-lens Sigma 105 mm 1:2.8), both for the roll and the two layers. The peeling of the two layers allows us to study the peeling process at the debonding region in more detail, and in addition to determining the Young's modulus of the backing of the tape (see Fig.\,S1, S2 and S3~\cite{supplemental}). Both set-ups are placed in a sealed box through which a mixture of compressed air and water vapor is flowing; varying the relative proportions allows maintaining a constant RH in the range $[1.9-98]\%$. The temperature is kept constant at 20$\pm 0.5~^\circ$C. The highest applied force of $1.27\,\mathrm{N}$ is chosen to avoid stick-slip effects: we focus on the steady-state regime of peeling for low peeling velocity ($V<10\,\upmu\mathrm{m}/\mathrm{s}$). It is customary to discuss the peeling speed as a function of the strain energy release rate $G$ which is directly linked to the applied load $F$ through the Rivlin equation~\cite{rivlin_effective_1997}, $G=F(1-\cos\theta)/b$, where $b=19\,\mathrm{mm}$ is the tape width and $\theta\sim\pi/2$ the peeling angle in our experiment (see Fig.\,S1). 

As usually found in the literature, we plotted in Fig.\,\ref{fig:peeling}(a) the energy release rate $G$ as a function of the velocity of peeling $V$ in a log log scale; note that we are able to determine the velocity of the tape roll down to $\sim\mathrm{nm}/\mathrm{s}$ so that it is not surprising that one doesn't observe the roll of tape on a desk unpeeling.  At a fixed RH, it is tempting to interpret our data with the  extensively used~\cite{barquins_tackiness_1981,crosby_adhesive_1999,villey_rate-dependent_2015} but hitherto unexplained power-law behavior of Maugis and Barquins~\cite{maugis_fracture_1978}, $G\propto V^n$. However, when looking at the effect of the RH, it would appear that the exponent $n$ strongly depends on RH, varying from $0.166\pm0.003$ at $\mathrm{RH}=1.9\%$ up to $0.507\pm0.031$ at $\mathrm{RH}=98\%$. This means that for small applied forces, a humid environment boosts the peeling velocity over four orders of magnitude. The strong dependence of the exponent on the RH is not completely understood. Moreover, a second mystery arises for these simple experiments: at low RH, we find an exponent close to the very small value $n \approx 1/8$ found by Barquins for a different adhesive tape~\cite{barquins_kinetics_1997}. In the following, we focus our discussion on understanding the effect of the humidity on the peeling and then try to understand this small exponent for low $\mathrm{RH}$.

\section{Discussion}

\begin{figure*}[th!]
    \centering
    \includegraphics[width=\textwidth]{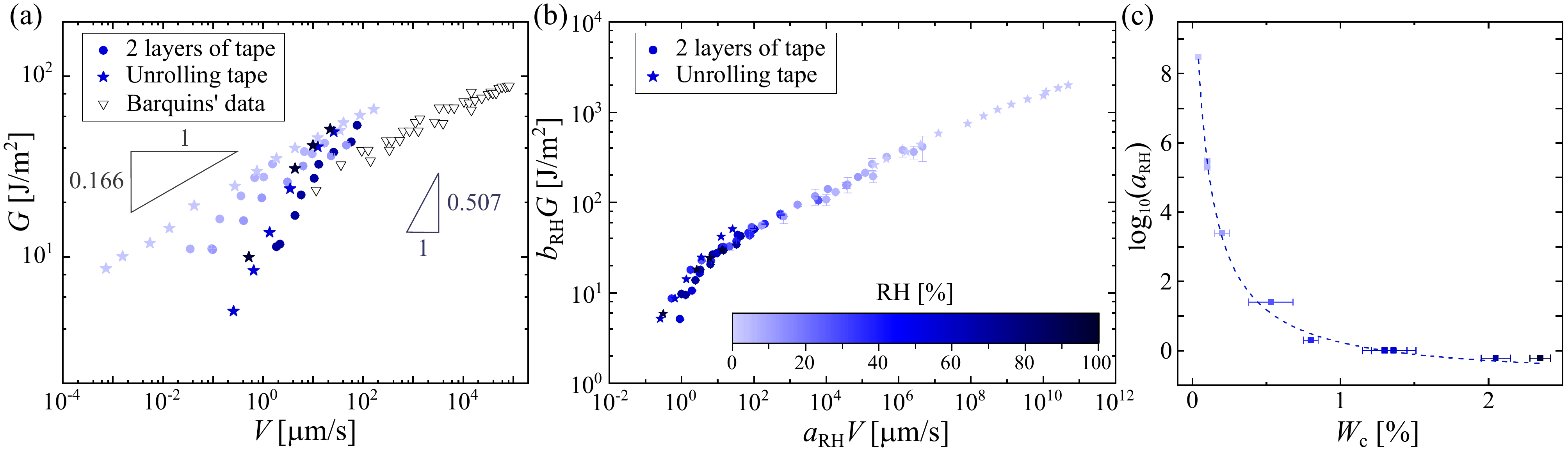}
    \caption{(a) Strain energy release rate $G$ as a function of peeling velocity $V$ of the adhesive for a suspended unrolling roll of tape (stars) and two layers of tape on glass (filled circles). The color scale (same as in (b)) indicates the different relative humidities. For a comparison, we plot Barquins' data~\cite{barquins_kinetics_1997} for the peeling of Scotch 3M 600. (b) Peeling master curve with $\mathrm{RH}_\mathrm{ref}=56.7\%$. The rescaled strain energy release rate $b_\mathrm{RH}\cdot G$ (with $b_\mathrm{RH} = \mathrm{RH}_\mathrm{ref}/\mathrm{RH}$) is plotted against the rescaled peeling velocity $a_\mathrm{RH}\cdot V$. The master curve is built with data reported in Fig.\,S4. (c) Logarithm of the rescaling factor $a_\mathrm{RH}$ as a function of the water content $W_\mathrm{c}$ in the adhesive tape. The dotted line is the best fit with Eq.\,\eqref{eq:analogie_RH}, following the concept of 'time-humidity' superposition. \footnotesize }
    \label{fig:peeling}
\end{figure*}

The minimal peeling energies measured in the present work at the low peeling velocities are about $4\,\mathrm{J}/\mathrm{m}^2$, which is still much higher than the Dupré interfacial work of adhesion for typical adhesive interfaces, which is around $0.1\,\mathrm{J}/\mathrm{m}^2$~\cite{creton_fracture_2016}. Thus, there must be a source of visco-elastic dissipation in the peeling dynamics. As the Young's modulus of the backing of the tape is found to be almost independent of RH, $\langle E_\mathrm{back} \rangle=1.44~\pm~0.26\,\mathrm{GPa}$ (see Fig.\,S3), such differences in the peeling of a dry and a humid environment must be due to important changes in the adhesive material itself. Contrary to the effect of humidity, the effect of temperature on the peeling of adhesives has been widely studied~\cite{kaelble_theory_1964,gent_adhesion_1969,derail_relationship_1998,marin_rheology_2006}. Gent \textit{et al.}~\cite{gent_adhesion_1969} and Kaelble~\cite{kaelble_theory_1964} were among the first to show that increasing the temperature leads to a smaller dissipation in the adhesive, and hence a larger peeling speed for a given force. They proposed a scaling procedure to collapse all temperature-dependent peeling curves onto a single master curve, following the same approach as the time-temperature superposition (TTS) principle in polymer rheology~\cite{ferry_viscoelastic_1961}. The peeling measurements for different temperatures are then rescaled in time by a coefficient $a_\mathrm{T}(T)$, comparable to the rheological factor in the TTS principle, to create a master curve that collapses around the data for a reference temperature $T_\mathrm{ref}$, usually chosen as the ambient temperature. The dependence of $a_\mathrm{T}$ on temperature follows the William-Landel-Ferry prediction for polymer liquids~\cite{marin_rheology_2006}. $a_\mathrm{T}$ reflects the molecular mobility of the polymer chains: the higher $a_\mathrm{T}$, the less mobile are the chains~\cite{ferry_viscoelastic_1961}. 

Our observations here are very similar, but as a function of humidity rather than temperature; we propose the construction of a peeling master curve for different relative humidities. Rescaling the peeling velocity by a factor $a_\mathrm{RH}(\mathrm{RH})$ and the strain energy release rate $G$ by $b_\mathrm{RH}=\mathrm{RH}_\mathrm{ref}/\mathrm{RH}$ with $\mathrm{RH}_\mathrm{ref}$ a reference relative humidity, we build the master curve with a reference data set for $\mathrm{RH}_\mathrm{ref}=56.7\%$, such that $a_\mathrm{RH}(56.7\%)=1$. This peeling master curve is plotted in Fig.\,\ref{fig:peeling}(b), based on  the data presented in Fig.\,S4. As in the construction of the rheology master curve~\cite{ferry_viscoelastic_1961}, the rescaling factor $a_\mathrm{RH}(\mathrm{RH})$ is manually tuned to obtain this peeling master curve (Fig.\,S5). We find that $a_\mathrm{RH}$ varies over 9 orders of magnitude, similarly as was reported in time-humidity rheology of different polymers~\cite{onogi_timehumidity_1962,zhou_confirmation_2001,fabre_time-temperature-water_2018}. $a_\mathrm{RH}$ has the same physical meaning as $a_\mathrm{T}$: a high value of $a_\mathrm{RH}$ means less mobile polymer chains in the adhesive. 

In order to connect the RH to the adhesion characteristics of the Scotch tape material, we apply the concept of 'time-humidity' superposition~\cite{ishisaka_examination_2004,hatzigrigoriou_temperature-humidity_2012,fabre_time-temperature-water_2018} to the dependence of $a_\mathrm{RH}$ with the water content $W_\mathrm{c}$ absorbed by the hygroscopic adhesive:
\begin{equation}
    \log_{10}(a_\mathrm{RH})=\frac{-D_1(W_\mathrm{c}-W_\mathrm{c,\,ref})}{D_2+W_\mathrm{c}-W_\mathrm{c,\,ref}}~,
\label{eq:analogie_RH}
\end{equation}
where $W_\mathrm{c,\,ref}$ is the water content of the adhesive at $\mathrm{RH}_\mathrm{ref}$ and $D_1$ and $D_2$ are empirical constants. This means that less water content leads to a high value of $a_\mathrm{RH}$: the polymer chains are less mobile when the amount of absorbed water is low. The amount of water is measured as a function of RH by a simple gravimetric test (Fig.\,S6). The dotted line in Fig.\,\ref{fig:peeling}(c) is the best fit to eq.\,\eqref{eq:analogie_RH}, with $D_1=0.87\pm0.05$,  $D_2=1.39\pm0.01\%$, and $W_\mathrm{c,\,ref}=1.30\%$ for $\mathrm{RH_{ref}}$ = 56.7\%. The values of $D_1$ and $D_2$ are in good agreement with known constants for different polymers~\cite{ishisaka_examination_2004,fabre_time-temperature-water_2018}. This shows that the rescaling factor $a_\mathrm{RH}$ of the peeling master curve is indeed linked to changes in the visco-elastic behavior of the adhesive. Following the analogy with the TTS principle, one possible mechanism to explain the dependence of $a_\mathrm{RH}$ on the water content in the PSA is the hydroplastization of the adhesive: when the water content increases, the glass transition temperature $T_\mathrm{g}$ of the adhesive decreases~\cite{hatzigrigoriou_diffusion_2011,hatzigrigoriou_temperature-humidity_2012}. Even though no direct measurement of the glass transition temperature was performed on our tape, such hydroplastization of acrylic adhesive has already been reported in the literature~\cite{bianchi_study_1990}: Bianchi \textit{et al.} reported a decrease of the $T_\mathrm{g}$ between 10 and $40^\circ$C. A similar trend is highly possible is the tape we used for our experiments. At low RH the glass transition temperature would be closer to the ambient temperature than at high RH. Thus, the polymer chains in the adhesive are less mobile in the adhesive at low humidity: the adhesive is closer to the glass transition at low RH and thus respond more elastically, and respectively, the viscous effects dominate at high RH.

Another way to estimate the visco-elasticity of the adhesive is to calculate the Deborah number: for high Deborah numbers, elasticity dominates, whereas viscous effects become important for small Deborah numbers. Following the analogy with the TTS principle~\cite{lakrout_influence_2001}, it is possible to estimate the Deborah number $\mathrm{De}$ for our peeling experiments:

\begin{equation}
  \mathrm{De}=\tau_\mathrm{d}a_\mathrm{RH}V/e
\end{equation}
with $\tau_\mathrm{d}$ the terminal relaxation time and $e$ the thickness of the adhesive layer. $\tau_\mathrm{d}$ is here estimated from values of the literature~\cite{lakrout_influence_2001} as no dynamic mechanical analysis were performed on the Scotch tape 810. By evaluating $\tau_\mathrm{D}\sim1000\,\mathrm{s}$, the Deborah number $\mathrm{De}$ would range in $[10^1-10^{12}]$: the adhesive is more elastic at low RH and more viscous at high RH. Note that a lower estimated value for the terminal relaxation time ($\tau_\mathrm{D}<100\,\mathrm{s}$) would even lead the Deborah number below 1, meaning the scotch tape would behave as a 'liquid-like' system and flow.

Finally, to support the robustness of our approach, we calculated the evolution of the velocity of peeling $V(t)$ for the unrolling of the tape followed over a month (Fig.\,\ref{fig:setup_illustration}) based on the measurement RH$(t)$ (see details in Supplementary Information). We show in Fig.\,\ref{fig:setup_illustration}(b) the calculated $V(t)$ indeed collapses with the experimental data. To summarize, the high impact of the relative humidity on the peeling of a Scotch tape is due to changes in the bulk visco-elastic properties of the adhesive. The hydroplastization of the adhesive then gives a satisfactory explanation for the dependence of the peeling velocity on the relative humidity: the higher the RH, the more easily the adhesive is peeled.


It is then tempting to attribute the very low exponent $n=0.166$ at $\mathrm{RH}=1.9\%$ presented in Fig.\,\ref{fig:peeling}(a) to the hydroplastization of the adhesive. However, this exponent is surprisingly close to the one found by Barquins ($n=0.146$)~\cite{barquins_kinetics_1997} where they did not monitor the RH for their experiments. The IR characterization of the tape 3M 600 they used presents no trace of water (see Fig.\,S7): the time-humidity superposition principle is not applicable for their experiments. Bulk viscoelastic dissipation is not the main peeling mechanism there. Chopin \textit{et al.}~\cite{villey_rate-dependent_2015,chopin_nonlinear_2018} recently proposed to explain such a low exponent by taking into account the non linear rheology of the stretched fibrils. Within the resolution of our experiments, the stretching of fibrils is rate independent (see Fig.S11): the adhesion curve $G(V)$ is dominated by the linear viscoelasticity of the adhesive and not the non linear rheology of fibrils. Here, we rather propose to take into account interfacial dissipation. These peeling experiments are close to the limit in which the adhesive behaves elastically. Therefore, in order to detach, a fracture has to propagate within the adhesive or between the adhesive and the backing. To rationalize the rate-dependence of the fracture energy of elastomers, Chaudhury \textit{et al.} proposed a fracture mechanism based on the kinetic theory of bond rupture~\cite{chaudhury_rate-dependent_1999,ghatak_interfacial_2000,hui_failure_2004}. According to Evans~\cite{evans_dynamic_1997}, when a polymer chain is stretched with a force $f$, its activation energy of dissociation decreases by $f\lambda$, where $\lambda$ is an activation length of a bond, usually approximated to the length of a chemical bond ($\lambda\sim0.1\,\mathrm{nm}$)~\cite{chaudhury_rate-dependent_1999}. This allows to quantify how the probability of failure of any bond in the polymer chain varies with the applied force. Chaudhury's model proposes the strain energy release rate $G$ to depend on the stretching velocity of a bond $V_\mathrm{stretch}$ as:

\begin{equation}
    G=\left( \frac{\Sigma_0}{2k_\mathrm{s}}\right)\left[ \left( \frac{k_\mathrm{B}T}{\lambda}\right)\ln \left( \frac{k_\mathrm{s}\lambda \tau_{-} V_\mathrm{stretch}}{n_\mathrm{bond} k_\mathrm{B}T} \right)\right]^2~,
    \label{eq:chaudhury_model}
\end{equation}
where $k_\mathrm{B}$ is Boltzmann constant, $T$ temperature, $n_\mathrm{bond}$ the number of bonds per polymer chain, ${\Sigma_0}$ the number of load-bearing polymer chains per unit area, $k_\mathrm{s}$ the stiffness of the polymer chain, and $\tau_{-}$ the characteristic time of bond dissociation. 

\begin{figure}[ht!]
    \centering
    \includegraphics[width=\columnwidth]{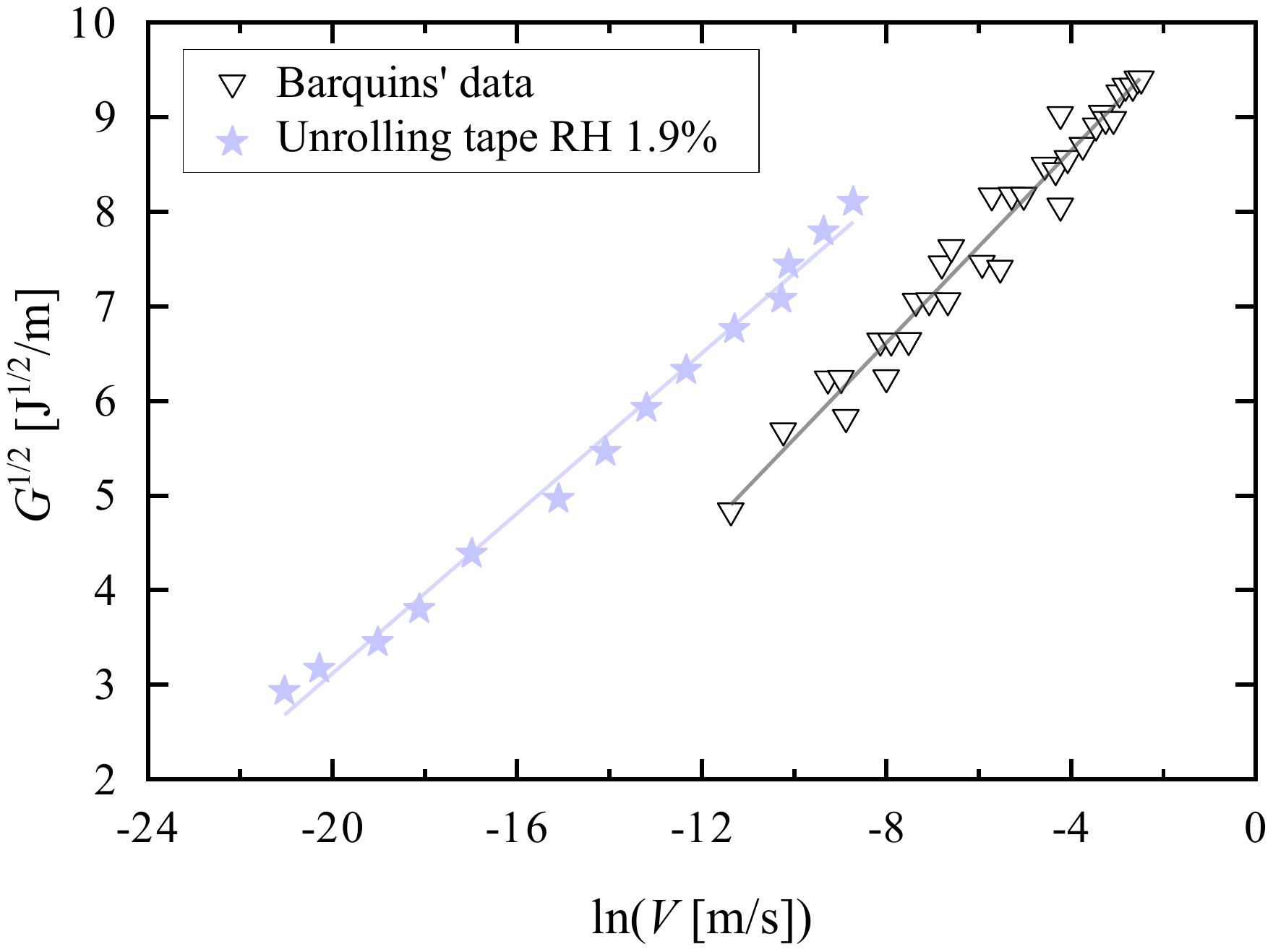}
    \caption{Square root of the strain energy release rate $G$ as a function of the logarithm of the peeling velocity $V$ for the unrolling tape roll at $\mathrm{RH}=1.9\%$ (stars) and Barquins' data~\cite{barquins_kinetics_1997} (triangle). For our experiments, we assume the stretching velocity of a bond $V_\mathrm{stretch}$ to be equal to the peeling velocity $V$. The solid lines are fits to Eq.\,\eqref{eq:chaudhury_model}. \footnotesize }
    \label{fig:chaudhury_model}
\end{figure}

In Fig.\,\ref{fig:chaudhury_model}, we plot $G^{1/2}$ as a function of $\mathrm{ln}(V)$ for our tape at $\mathrm{RH}=1.9\%$ and for Barquins' data~\cite{barquins_kinetics_1997}. Chaudhury's model describes both data sets very well, explaining the small exponent: the behavior is in fact not a power-law behavior with a small exponent, but rather a logarithmic dependence due to the presence of an activated process.

From the fits, we can then obtain estimates of the spring constant $k_\mathrm{s}$ and the characteristic time $\tau_{-}$. Assuming $n_\mathrm{bond} \in[100-1000]$~\cite{chaudhury_rate-dependent_1999,hui_failure_2004,zhu_rate_2019}, and $\Sigma_0\sim 10^8\,\mathrm{chains}/\mathrm{m}^2$~\cite{chaudhury_rate-dependent_1999,macron_equilibrium_2018,zhu_rate_2019}, we find respectively $k_\mathrm{s}=11.6\pm0.6\,\mathrm{mN}/\mathrm{m}$ and $\tau_{-}$ between $3.4\times10^4$ and $3.4\times10^5$ s for our data and $k_\mathrm{s}=8.0\pm0.6\,\mathrm{mN}/\mathrm{m}$ and $\tau_{-}$ between $6.6\times10^2$ and $6.6\times10^3$ s for Barquins' data . Note that these stiffnesses are one order of magnitude lower than the typical stiffness of the polymer chain for a strong bond ($k_\mathrm{s}=0.5\,\mathrm{N}/\mathrm{m}$), meaning that the bond in the adhesive breaks long before its full extension~\cite{cedano-serrano_molecular_2019}. Furthermore, according to Eyring's model, the bond dissociation time $\tau_{-}$ is:

\begin{equation}
    \tau_{-}=\frac{h}{k_\mathrm{B}T}\exp\left(\frac{E_\mathrm{a}}{k_\mathrm{B}T}\right)~,
\end{equation}
where $h$ is Planck's constant. Thus, the activation energy of bond dissociation $E_\mathrm{a}=105\pm3\,\mathrm{kJ}/\mathrm{mol}$ in our experiment and $E_\mathrm{a}=91\pm3\,\mathrm{kJ}/\mathrm{mol}$ for Barquins'. These energies are smaller than the dissociation energy of a covalent bond ($\sim400\,\mathrm{kJ}/\mathrm{mol}$)~\cite{cedano-serrano_molecular_2019} and the decomposition activation energy of an acrylic adhesive ($\sim200\,\mathrm{kJ}/\mathrm{mol}$)~\cite{wang_thermal_2016}. However, they are of the order of respectively 7 and 6 hydrogen bond dissociation ($E_\mathrm{H\,bound}\sim15\,\mathrm{kJ}/\mathrm{mol}$~\cite{cedano-serrano_molecular_2019,macron_equilibrium_2018}). Interestingly, the small differences between the two data sets can be attributed to the difference of the adhesives in the two different types of Scotch tape, which might be a useful avenue to pursue for improving pressure sensitive adhesives. 

\begin{figure}[h!]
    \centering
    \includegraphics[width=\columnwidth]{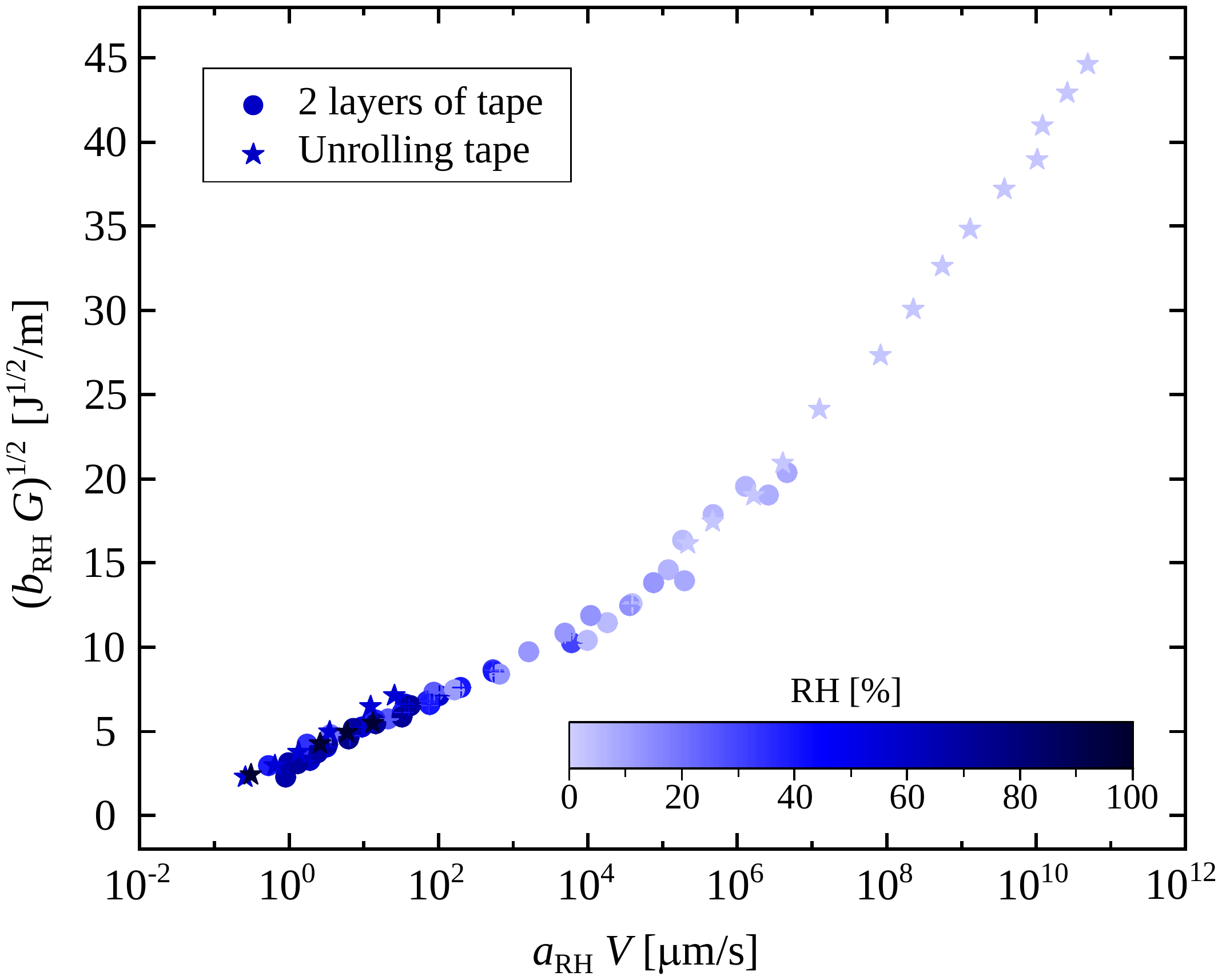}
    \caption{Square root of the rescaled strain energy release rate $b_\mathrm{RH}G$ as a function of the rescaled peeling velocity $a_\mathrm{RH}V$ in a log-lin scale. This graph presents the same data as in Fig. 2(b) but plotted to compare it to Chaudhury's theory. \footnotesize}
    \label{fig:SI_chaudhury_master_curve}
\end{figure}

Such a fracture-like theory does not preclude any influence of visco-elasticity on the fracture behavior. For instance, viscoelastic fracture that agrees with Griffith fracture theory has been observed on much softer materials than the PSA we used~\cite{bonn_delayed_1998}. Then arises the question of how to link Chaudhury's theory and the building of a peeling master curve. In Fig.\,\ref{fig:SI_chaudhury_master_curve}, the data presented in the peeling master curve (Fig. 2(b)) are plotted to be compared with Chaudhury's model (see Eq. 2): the square root of the rescaled strain energy release rate $b_\mathrm{RH}G$ as a function of the rescaled peeling velocity $a_\mathrm{RH}V$ in a log-lin scale.

As the data follow two lines, we distinguish two regimes. We attribute the regime at 'high' humidity to a competition between the interfacial dissipation, described by Chaudhury's model, and the viscous dissipation in the adhesive. Indeed, such bulk dissipation are not taken into account in Chaudhury's model. To our knowledge, there is no complete theory that proposes a picture of the whole mechanism. Note the limit between these two behaviors is not sharp as part of the data at $\mathrm{RH}=7.4\%$ follows the same trend as the ones at $\mathrm{RH}=1.7\%$. More extensive experiments with a controlled chemistry of the adhesive would be needed to clearly distinguish the boundary between bulk dissipation, reflected with the construction of the peeling master curve, and the interfacial ones, explained by Chaudhury's model. Probe tests might be used to quantify the adhesion and investigate further these bulk and interfacial effects, as recently proposed by Wang \textit{et al.}~\cite{wang_bulk_2020}. Additionally, extra theoretical efforts are desired to construct a model that can include both the bulk and interfacial dissipation.


\section{Conclusion}

In sum, we provided an understanding of the force necessary to peel a typical PSA depending on the environmental conditions. The first surprise is the sensitivity to the humidity: the more humid the environment, the faster the tape peels for a given force, with a power-law relation between speed and force that can be successfully described by a master curve applying 'time-humidity' superposition. The second surprise is that for very low humidities, the adhesive becomes strongly elastic and a completely different regime emerges. Chaudhury's theory of rate-dependent bond fracture allows to quantitatively describe the observed logarithmic dependence of the peeling velocity on the strain energy release rate as an activated process, explaining also previously reported power-law behavior with an inexplicably small exponent in this regime. This work opens the way to develop a complete quantitative understanding of soft visco-elastic adhesives by showing that, depending on external parameters either viscous or elastic behavior can be expected and explained.  

\section*{Acknowledgements}
This project has received funding from the European Research Council (ERC) under the European Union's Horizon 2020 research and innovation program (Grant agreement No. 833240). We are much indebted to M. Golden who first suggested this experiment, and acknowledge very helpful discussions with  M. Ciccotti and C. Creton.

\bibliographystyle{apsrev4-1}
\bibliography{PSA_humidity}

\end{document}


\author{Marion Grzelka}
\email{m.grzelka@uva.nl}
\affiliation{Van der Waals-Zeeman Institute, Institute of Physics, University of Amsterdam, 1098XH Amsterdam, The Netherlands}
\author{Stefan Kooij}
\affiliation{Van der Waals-Zeeman Institute, Institute of Physics, University of Amsterdam, 1098XH Amsterdam, The Netherlands}
\author{Sander Woutersen}
\affiliation{Van 't Hoff Institute for Molecular Science, University of Amsterdam, 1098XH Amsterdam, The Netherlands}
\author{Mokhtar Adda-Bedia}
\affiliation{Université de Lyon, Ecole Normale Supérieure de Lyon, Université Claude Bernard, CNRS, Laboratoire de Physique, Lyon F-69342, France}
\author{Daniel Bonn}
\affiliation{Van der Waals-Zeeman Institute, Institute of Physics, University of Amsterdam, 1098XH Amsterdam, The Netherlands}


\title  {Supplementary Information - Transition from viscoelastic to fracture-like peeling of pressure-sensitive adhesives}



\maketitle
\begin{singlespace}
\clearpage
\section{Young's modulus of the backing of the tape}

\begin{figure*}[ht!]
    \centering
    \includegraphics[width=\textwidth]{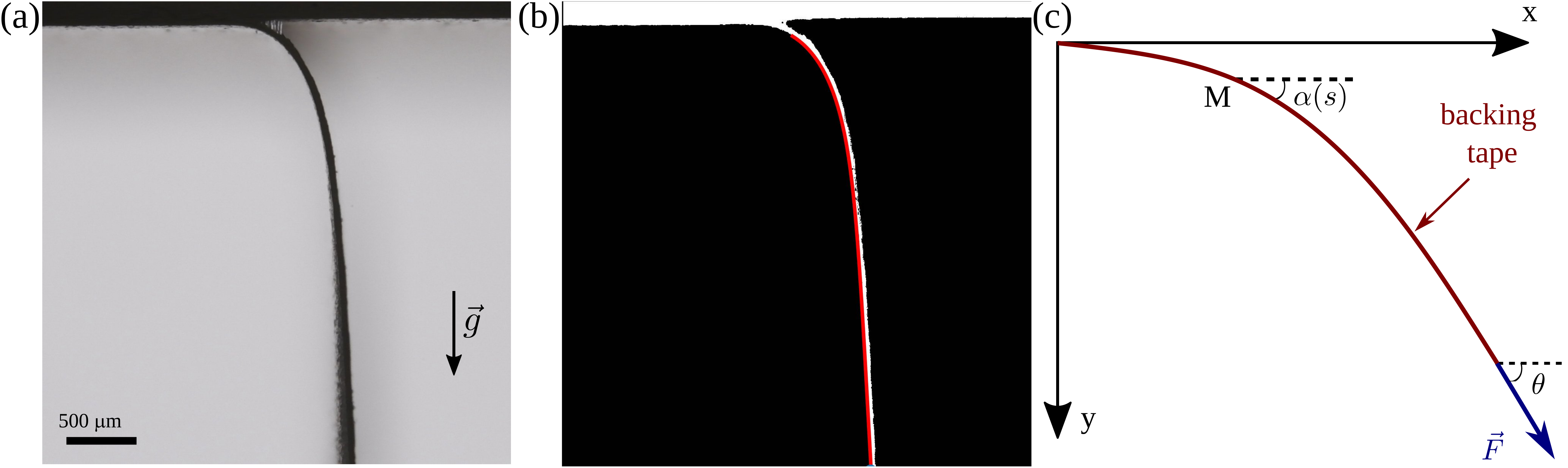}
    \caption{Procedure to extract the backing profile of the tape. (a) Picture of the peeling of the Scotch tape (3M 810  Magic Scotch) at $\mathrm{RH}=12.5\%$ with a load $F=803\,\mathrm{mN}$. (b) Binarization of the picture to extract the backing profile (red). (c) Illustration of the peeling. A load  $\vec{F}$ is applied to the tape with a peeling angle $\theta$. $\alpha(s)$ is the local angle in M. \footnotesize}
    \label{fig:SI_curvature}
\end{figure*}

By peeling two layers of the adhesive over each other on an upside-down glass plate, we can track the peeling front but also the shape of the backing of the tape (Fig.\,S\ref{fig:SI_curvature} (a) and (b)). From the pictures, we extract the local angle $\alpha$ (Fig.\,S\ref{fig:SI_curvature}(c)) as a function of the curvilinear abscissa $s$ defined as:
\begin{equation}
    ds \overrightarrow{t}=ds\cos(\alpha)\overrightarrow{e_x}+ds\sin(\alpha)\overrightarrow{e_y}
\end{equation}

Using simple elastica model for an inextensible beam~\cite{love_treatise_1944}, the local angle $\alpha(s)$ is:
\begin{equation}
    \alpha(s)=\theta-4\arctan\left(\tan\left(\frac{\theta}{4}\right)\exp\left(-\frac{s-s_0}{R_\mathrm{c}}\right)\right)
\end{equation}
with $\theta$ the peeling angle, $s_0$ the abscissa for the clamping condition and $R_\mathrm{c}$ the radius of curvature of the backing of the tape. Moreover the radius of curvature is directly linked to the geometry of the backing of the tape and its Young's modulus $E_\mathrm{back}$:
\begin{equation}
    R_\mathrm{c}=\sqrt{\frac{E_\mathrm{back}be^3}{12F(1-\cos\theta)}}
    \label{eq:young}
\end{equation}
with $b=19\,\mathrm{mm}$ the width of the tape and $e=38\,\upmu$m the thickness of the backing of the tape. 

\begin{figure*}[ht!]
    \centering
    \includegraphics[height=8cm]{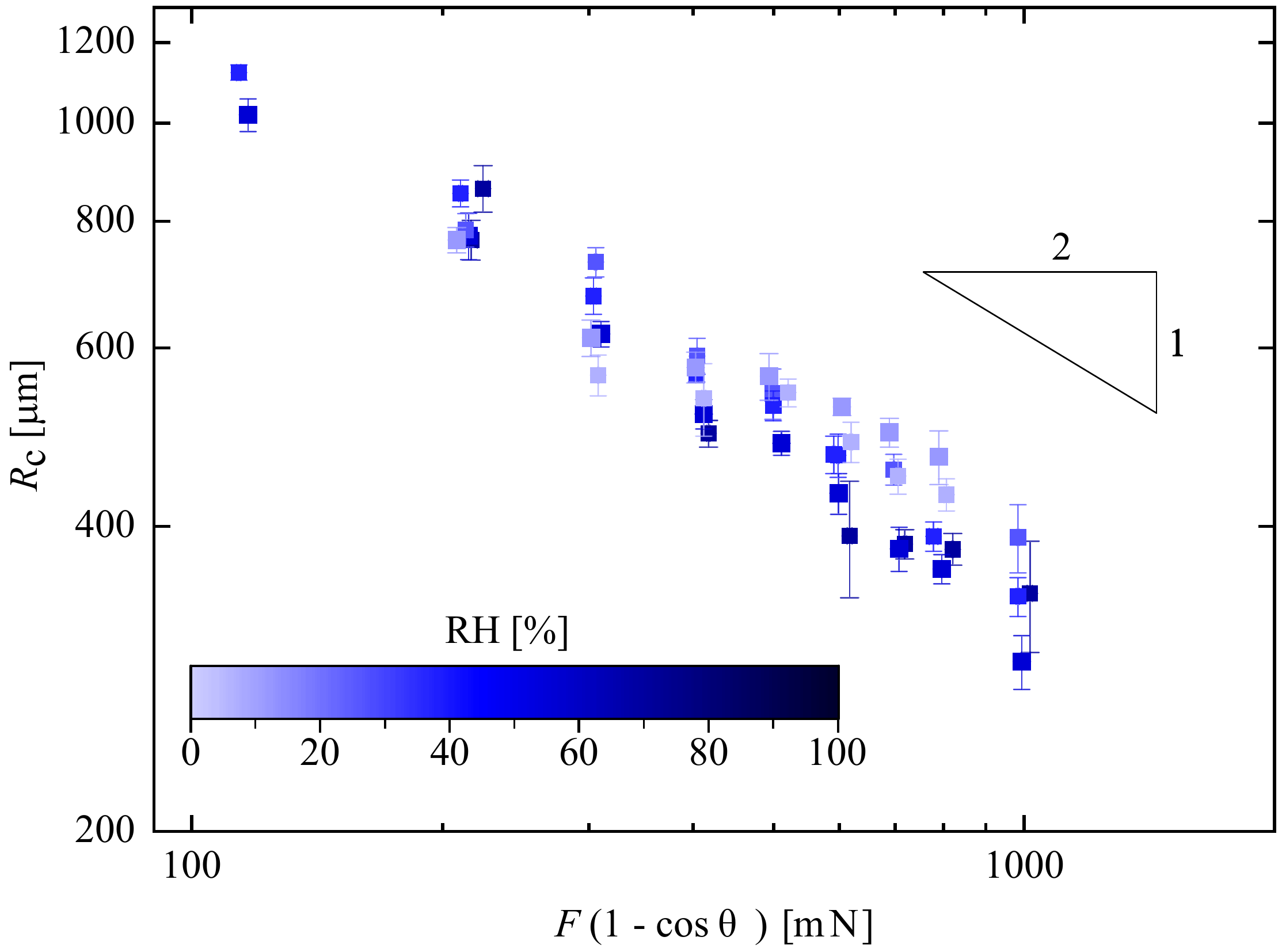}
    \caption{Radius of curvature $R_\mathrm{c}$ of the backing of the PSA as a function of the applied force $F(1-\cos\theta)$. The color scale indicates the different relative humidity RH.  \footnotesize}
    \label{fig:SI_Rc vs force}
\end{figure*}

In Fig.\,S\ref{fig:SI_Rc vs force}, the extracted radius of curvature $R_\mathrm{c}$ is represented as a function of the applied load $F (1-\cos\theta)$ for different humidities. The radius of curvature $R_\mathrm{c}$ follows the equation \eqref{eq:young}, scaling as the inverse of the square root of the applied load.

\begin{figure*}[ht!]
    \centering
    \includegraphics[height=8cm]{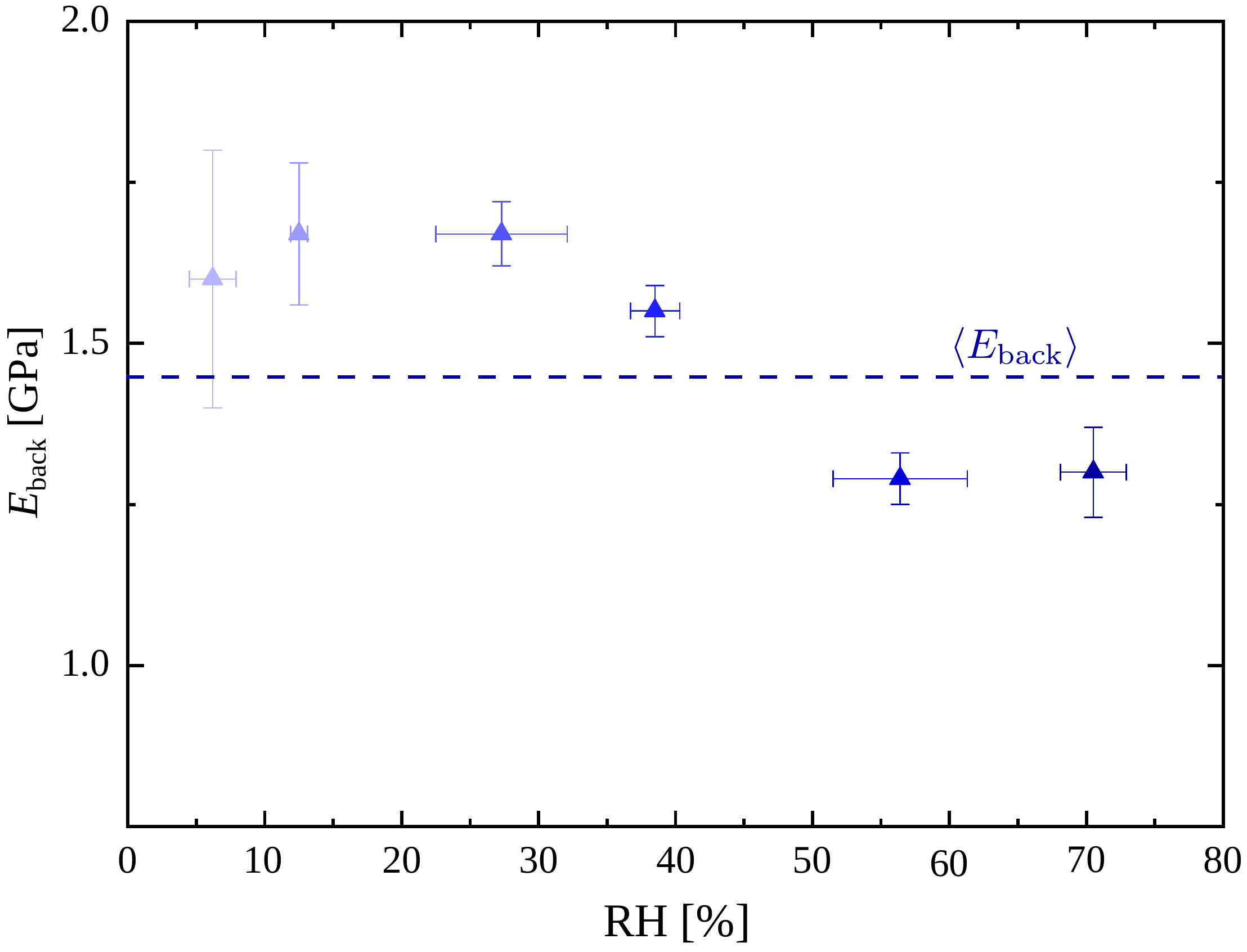}
    \caption{Young's modulus $E_\mathrm{back}$ of the backing of the PSA as a function of the relative humidity RH. The dotted line represents the mean value $\langle E_\mathrm{back}\rangle=1.44\,\mathrm{GPa}$. \footnotesize}
    \label{fig:SI_young_modulus}
\end{figure*}

The Young's modulus $E_\mathrm{back}$ of the backing of the tape for different humidities can be extracted as shown in Fig.S\,\ref{fig:SI_young_modulus}. The Young's modulus is almost independent of the relative humidity: we find $\langle E_\mathrm{back} \rangle=1.44\pm0.26\,\mathrm{GPa}$, in good agreement with values from the literature~\cite{molinari_peeling_2008,kovalchick_mechanics_2011}. We conclude that the impact of the relative humidity on the peeling velocity is due to change in the adhesive and not in the properties of the backing of the tape.

\clearpage
\section{Peeling experiments at different relative humidities}

\begin{figure*}[h!]
    \centering
    \includegraphics[height=7cm]{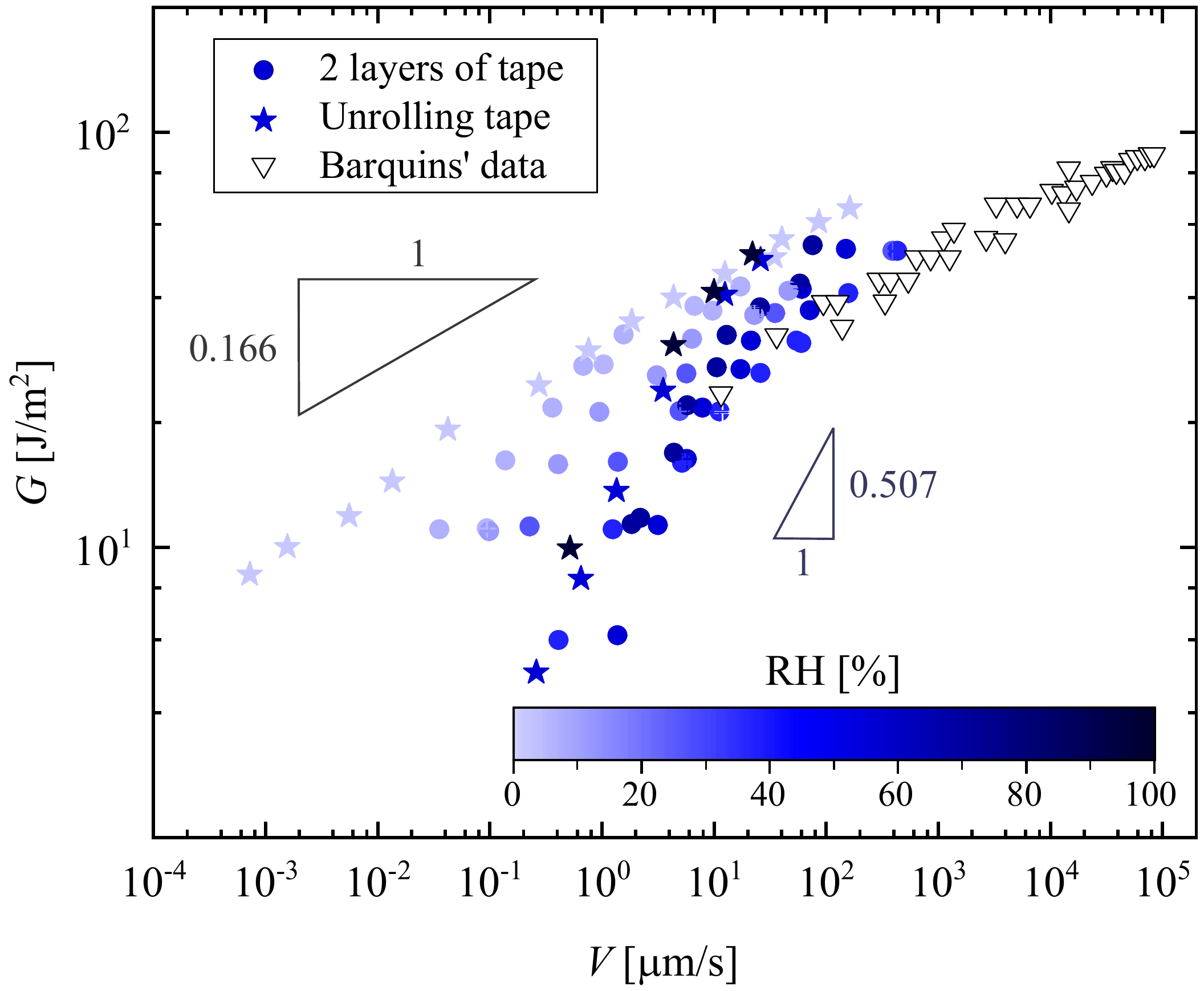}
    \caption{Strain energy release rate $G$ as a function of the peeling velocity $V$ of the adhesive for the unrolling tape and the two layers of the adhesive. The color map indicates the different relative humidity RH. We also plot Barquins' data~\cite{barquins_kinetics_1997} for the subcritical crack growth mode. \footnotesize}
    \label{fig:SI_peeling_all_data}
\end{figure*}

For the sake of clarity, only part of the peeling data are presented in Fig.\,2(a). In Fig.\, S\ref{fig:SI_peeling_all_data}, we present  all the data used to construct the peeling master curve presented in Fig.\,2(b).

\begin{figure*}[h!]

    \centering
    \includegraphics[height=7cm]{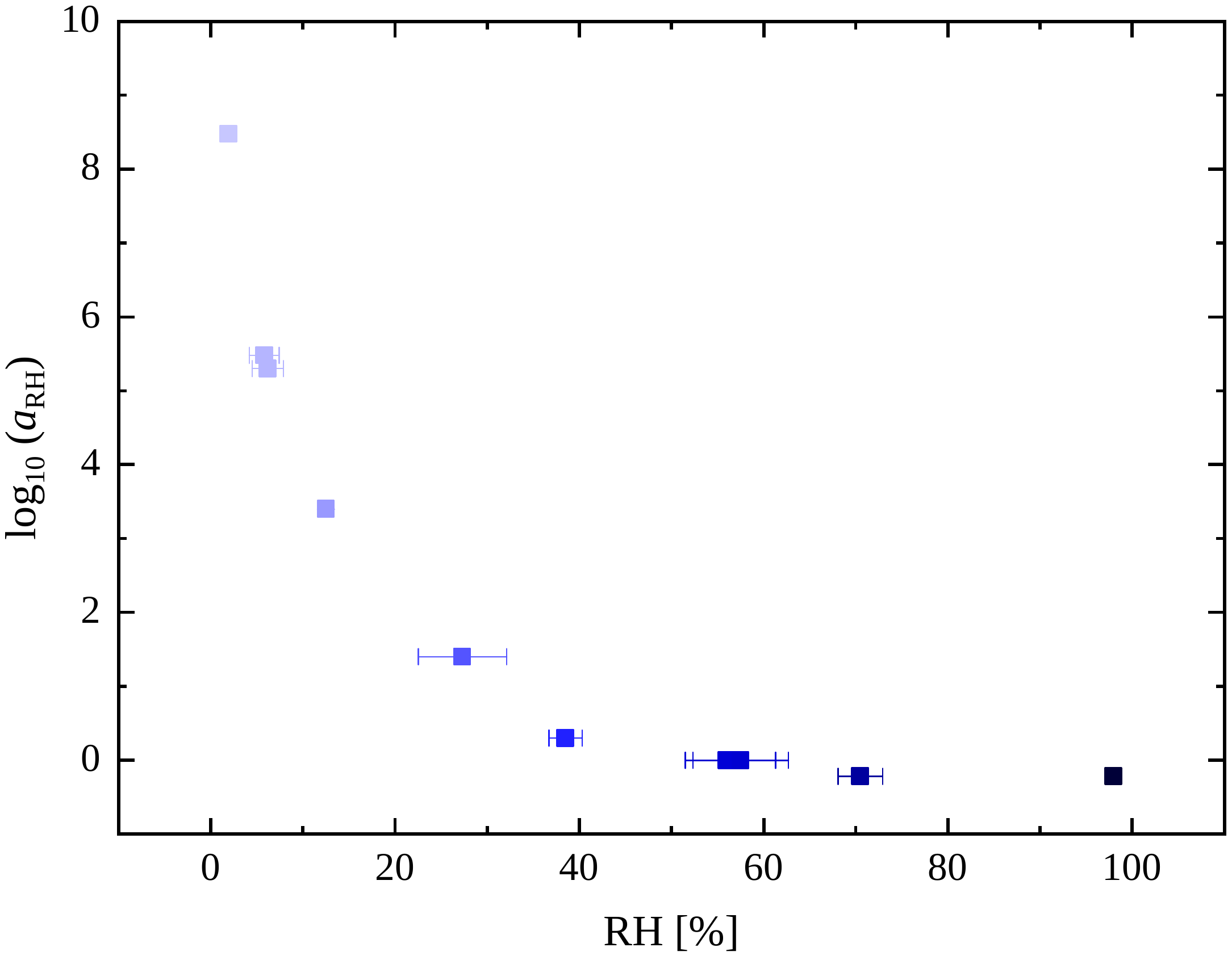}
    \caption{Rescaling factor $a_\mathrm{RH}$ that allows for the collapse of the data, i.e. to construct the master curve in Fig. 2(b) as a function of the relative humidity $\mathrm{RH}$.\footnotesize}

    \label{fig:aRH vs RH}
\end{figure*}

The peeling master curve (Fig.\,2(b)) is obtained by tuning the rescaling factor $a_\mathrm{RH}(\mathrm{RH})$. In Fig.\,S\ref{fig:aRH vs RH}, $a_\mathrm{RH}$ is shown to decrease with $\mathrm{RH}$.

\clearpage
\section{Absorption of water by the adhesive layer}
In order to connect the RH to the adhesion characteristics of the Scotch tape material, we measure the amount of water absorbed by the hygroscopic adhesive as a function of RH by using a simple gravimetric test. Samples of tape of one meter long are dried in a sealed box under a nitrogen flow at ambient temperature for several days. The dried samples are then exposed to different relative humidity for typically 2 days, to ensure the saturation of absorption of water. The water content for each relative humidity is deduced by weighing the samples (Mettler Toledo MS205DU, accuracy $0.01\,\mathrm{mg}$). The percentage of water content, $W_\mathrm{c}$, is determined by $W_\mathrm{c}=\frac{M_\mathrm{f}-M_\mathrm{0}}{M_\mathrm{0}}\times 100$,
where $M_\mathrm{f}$ and $M_\mathrm{0}$ are the weights of wet and dried specimens, respectively.

\begin{figure*}[h!]

    \centering
    \includegraphics[height=7cm]{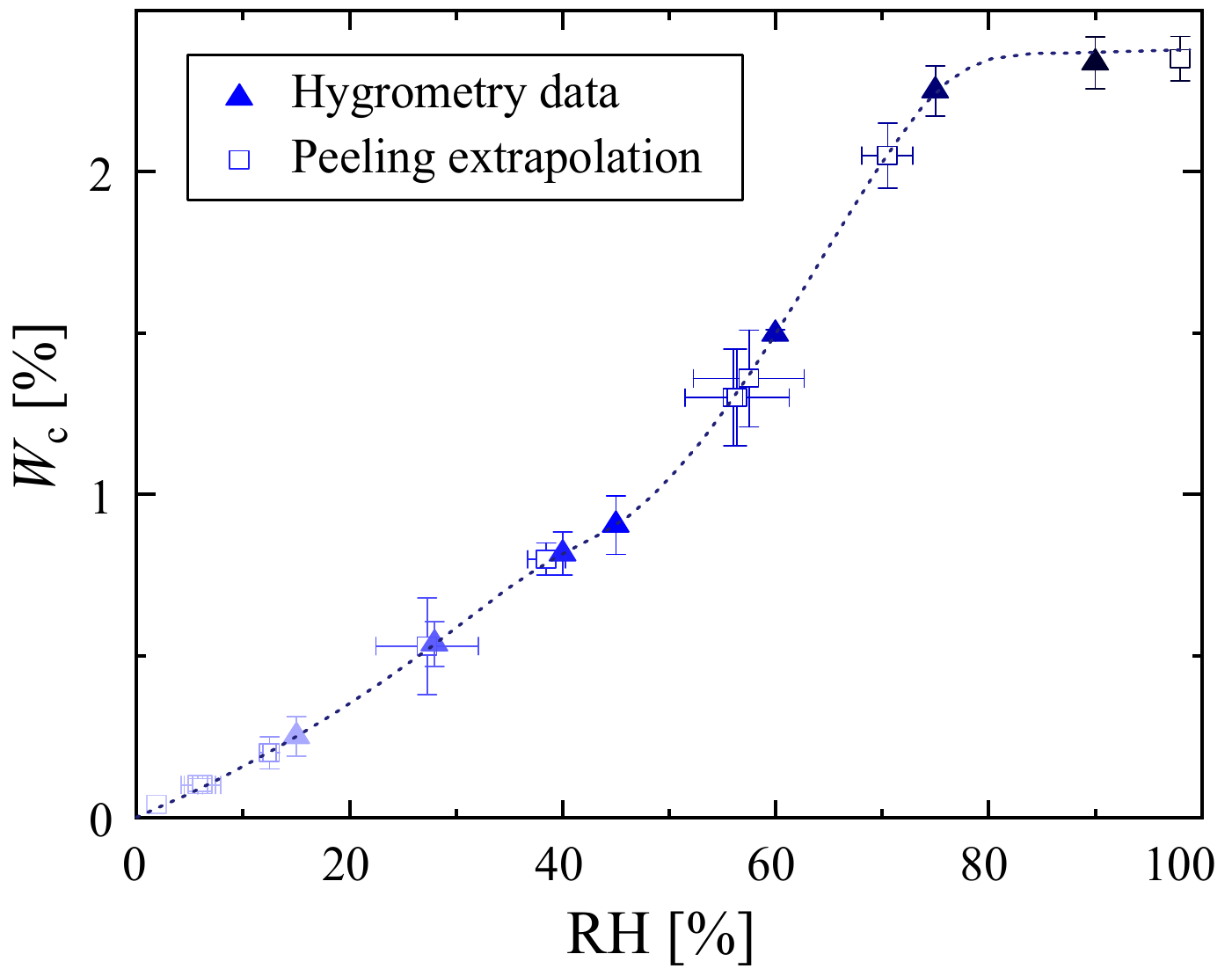}
    \caption{Water content in the adhesive $W_\mathrm{c}$ measured by gravimetric test (triangle) as a function of the relative humidity RH. The dotted line is a spline interpolation used to extrapolate the water content corresponding to the peeling experiments (square).\footnotesize}
    \label{fig:SI_water_content}
\end{figure*}

In Fig.\,S\ref{fig:SI_water_content}, the water uptake is shown to increase non-linearly with the relative humidity. At high humidity, $W_\mathrm{c}$ reaches more than $2.35\%$. We used a spline interpolation (dashed line) on the hygrometry data to extrapolate the water content corresponding to  the peeling experiments (square symbols).

\clearpage
\section{IR spectrophotometry}
\subsection{Comparison between tapes 3M 810 and 600}
\begin{figure*}[h!]
    \centering
    \includegraphics[width=\textwidth]{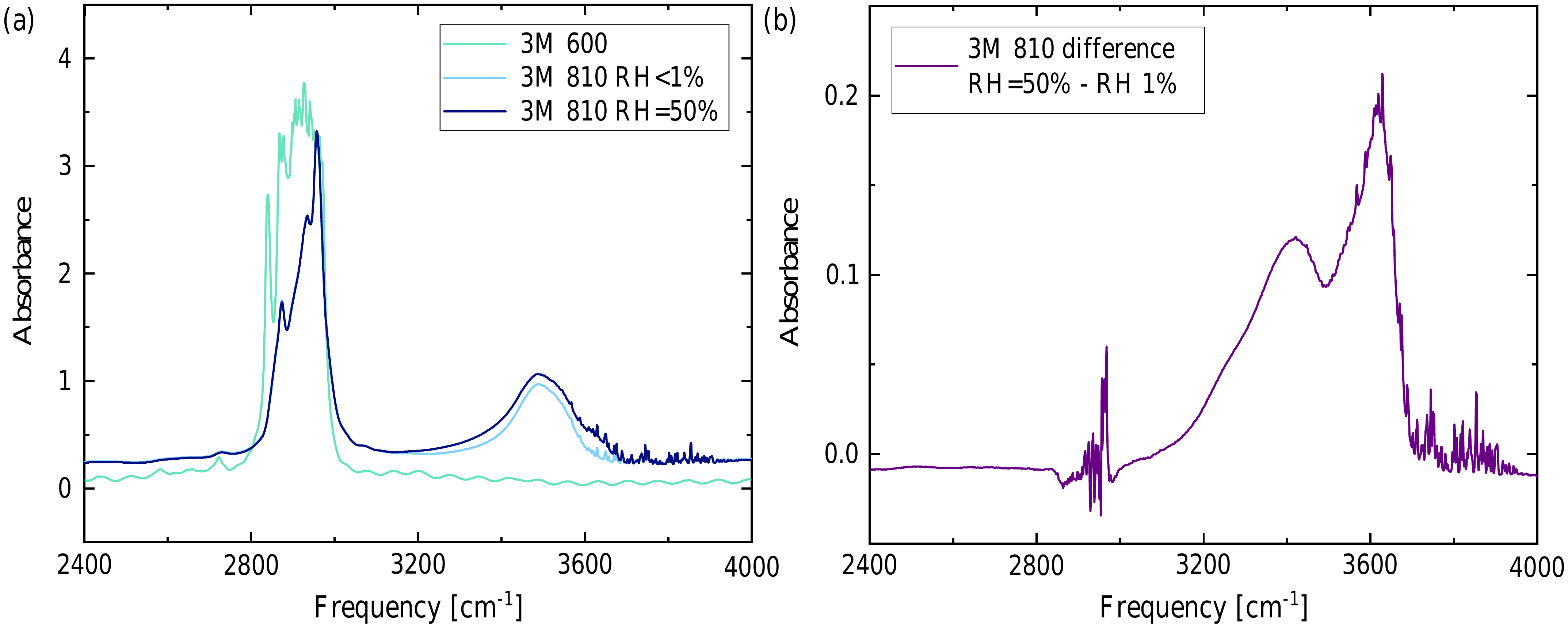}
    \caption{(a) IR-spectra of the tape 3M 810 at ambient RH (RH$\sim 50\%$) and dried (RH$< 1\%$) and the tape 3M 600 at ambient RH. (b) Difference of the IR-spectra for the wet and dried tape 3M 810.\footnotesize}
    \label{fig:SI_IR_analysis}
\end{figure*}

A Bruker Vertex 70 FTIR spectrometer (Bruker, Billerica, MA) is used to measure the 1D-IR (FTIR). The absorption spectra were recorded at a wavenumber resolution of $2\,\mathrm{cm}^{-1}$. We averaged 32 scans for every spectrum. We compare three samples: the tape 3M 810 at ambient RH (RH$\sim 50\%$), the same tape dried for 5h under a nitrogen flow (RH$< 1\%$) and the tape 3M 600 at ambient RH, used by Barquins \textit{et al.}~\cite{barquins_kinetics_1997}. Fig.\,S\ref{fig:SI_IR_analysis}(a) compares the spectra of the three samples.

Scotch tape 3M 810 presents a broad peak around $3400\,\mathrm{cm}^{-1}$ due to OH stretching for both the dried and the ambient RH samples. We still observe the OH peak after 5h of drying, indicating this remaining peak must be due to OH side groups in the tape.  Fig.\,S.\ref{fig:SI_IR_analysis}(b) presents the difference spectrum between the ambient RH and the dried samples: the double OH-peak agrees with the spectrum of H$_2$O in organic environment~\cite{troshin_effect_2008}. This clearly indicates the presence of water in the adhesive at ambient RH and confirms that the mass uptake we measured by gravimetric test is indeed water uptake.

The IR-spectrum of Scotch tape 3M 600 reveals the different chemistry of this tape compared to 3M 810 as there is no peak around $3400\,\mathrm{cm}^{-1}$, indicating the complete absence of water and OH side groups. This explains why Barquins \textit{et al.} did not report any effect of the humidity on the peeling of this tape.

\clearpage
\subsection{Drying of the tape 3M 810}

\textcolor{black}{On top of the mechanical changes in the tape, water may alter the interactions between the polymer chains in the adhesive~\cite{huacuja-sanchez_water_2016}. In order to investigate this effect, we dried the tape 3M 810 in the spectrometer, taking IR spectra each 2 min as described above.} 

\begin{figure*}[h!]
    \centering
    \includegraphics[width=\textwidth]{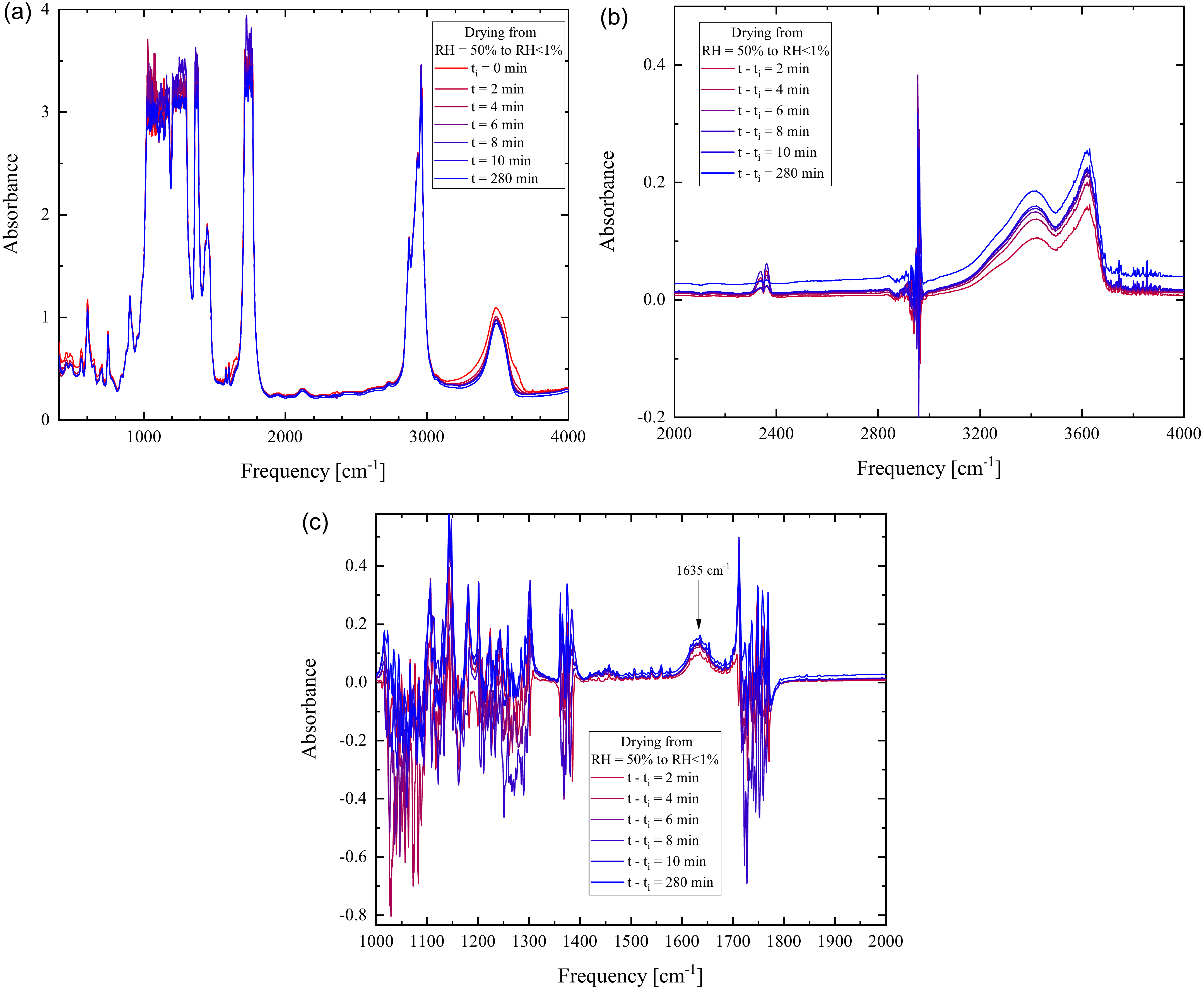}
    \caption{\textcolor{black}{(a) IR-spectra of the tape 3M 810 at ambient RH (RH$\sim 50\%$, $t_\mathrm{i}=0\,$min) and drying down to RH$< 1\%$ at $t=240\,$min. (b) and (c) Difference of the IR-spectra for the wet and dried tape 3M 810 at different drying time.\footnotesize}}
    \label{fig:SI_IR_drying}
\end{figure*}

\textcolor{black}{As shown in Fig.\,S\ref{fig:SI_IR_drying}(a), the IR-spectra of the drying tape differ systematically with the drying time for two specific peaks, respectively around $3400\,\mathrm{cm}^{-1}$ (see Fig.\,S\ref{fig:SI_IR_drying}(b)) and $1635\,\mathrm{cm}^{-1}$ (see Fig.\,S\ref{fig:SI_IR_drying}(c)). The peak at $3400\,\mathrm{cm}^{-1}$ has been described above as the one of H$_2$O in organic environment~\cite{troshin_effect_2008}. The peak around $1635\,\mathrm{cm}^{-1}$ is due to H$_2$O scissors bending vibration of water~\cite{celino_qualitative_2014}. As these changes in the IR-spectra are both due to water, the IR spectra indicates that other secondary interactions in the polymer chains are negligible.}

\section{Calculation of the velocity of peeling from relative humidity measurements}

Based on the time-humidity superposition principle and the peeling master curve, it is possible to calculate the evolution of the velocity of peeling $V(t)$ of the tape in time knowing the evolution of the relative humidity. Here we detail how we proceeded to calculate $V(t)$ in Fig. 1(b) (red line) based on the measurements of RH$(t)$.\\

The peeling master curve presented in Fig.~2(b) is fitted around the range of the data for the peeling experiment over a month: $b_\mathrm{RH}G_\mathrm{1\, month}\in [10^0-10^6]\,\mathrm{J}/\mathrm{m}^2$. This part of the data can be fitted as double power law:
\[
   b_\mathrm{RH}G = 
\begin{dcases}
    \beta(a_\mathrm{RH}V)^{\alpha_1},& \text{if } \mathrm{RH}\geq \mathrm{RH}_\mathrm{cut}\\
    \beta(a_\mathrm{RH}V)^{\alpha_2}(b_\mathrm{RH}(\mathrm{RH}_\mathrm{cut})G/\beta)^{\frac{\alpha_1-\alpha_2}{\alpha_1}},              & \text{otherwise}
\end{dcases}
\]

\begin{figure*}[h!]
    \centering
    \includegraphics[height=7cm]{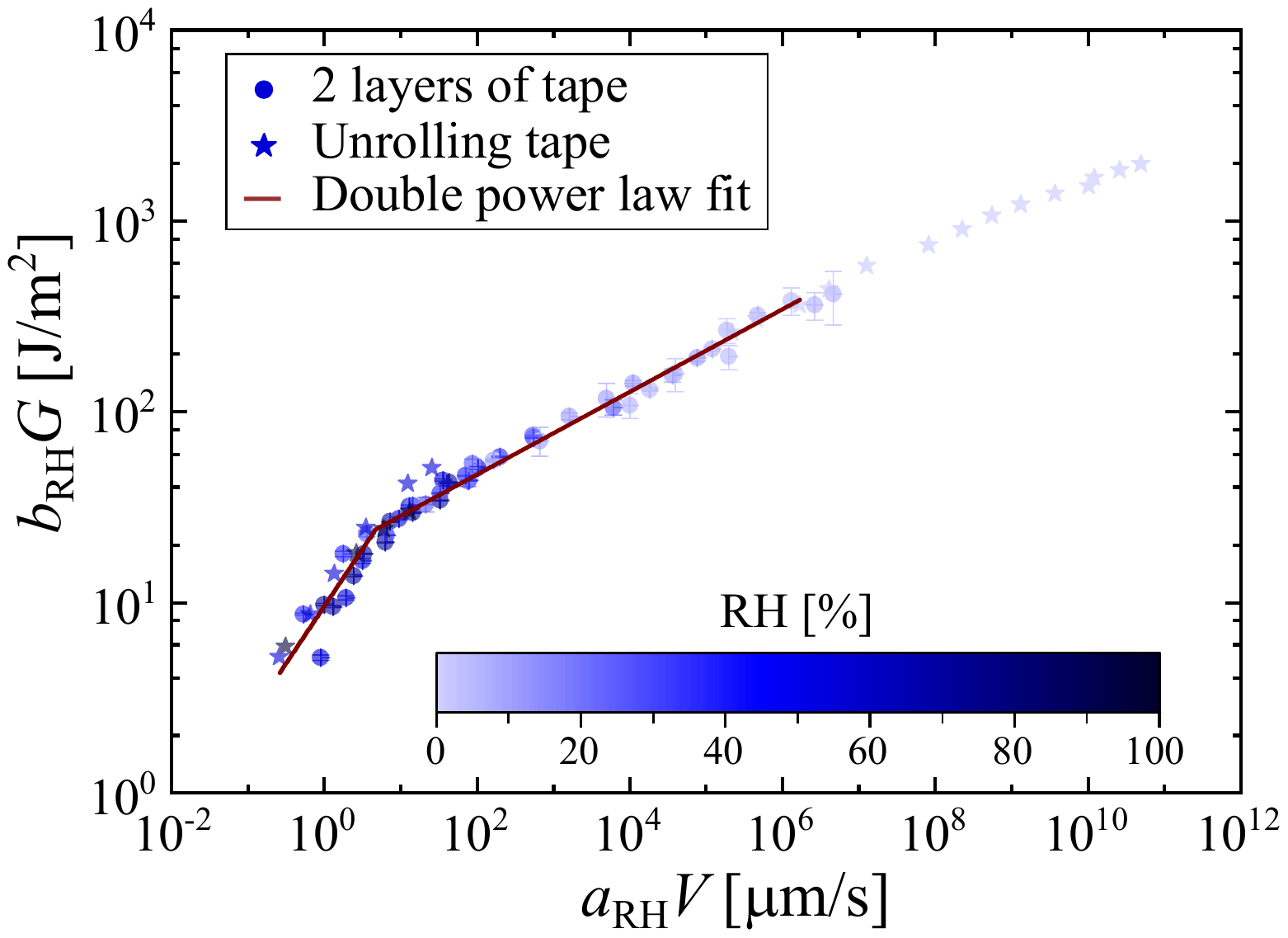}
    \caption{Double power law fit of the peeling master curve with $\mathrm{RH}_\mathrm{ref}=56.7\%$. The rescaled strain energy release rate $b_\mathrm{RH}\cdot G$ (with $b_\mathrm{RH} = \mathrm{RH}_\mathrm{ref}/\mathrm{RH}$) is plotted against the rescaled peeling velocity $a_\mathrm{RH}\cdot V$. The red line is the best double power law fit.  \footnotesize}
    \label{fig:SI_fit_master_curve}
\end{figure*}

Figure\,S\ref{fig:SI_fit_master_curve} presents the best double power law fit (red line), with $\mathrm{RH}_\mathrm{cut}=25.3\%$, $\alpha_1~=~0.60\pm0.23$, $\alpha_2=0.216\pm0.005$ and $\beta=9.73\pm3.43\,10^{-6\alpha_1}\mathrm{J}^{\alpha_1}\mathrm{s}^{\alpha_1}/\mathrm{m}^{\alpha_1}$. 

\begin{figure*}[h!]
    \centering
    \includegraphics[height=7cm]{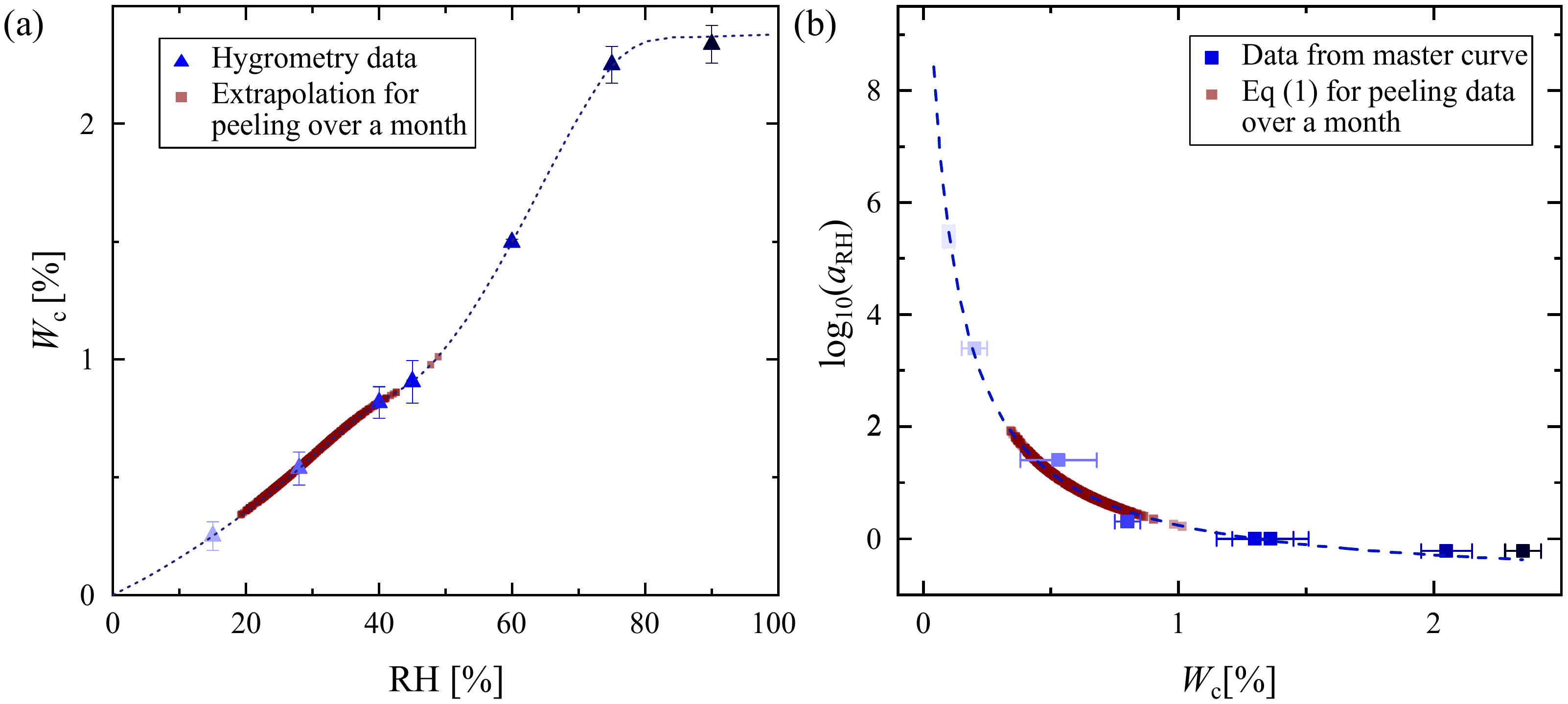}
    \caption{(a) Extrapolation of the water uptake for the peeling experiment over a month (red) based on the hygroscopic data (blue) with the relative humidity data from Fig.~1(b). (b) Calculation of the rescaling factor $a_\mathrm{RH}$ for the peeling experiment over a month (red) with the equation (1). Blue points are the data from the peeling experiments with different forces, presented in Fig. 2(c).  \footnotesize}
    \label{fig:SI_aRH_over_a_month}
\end{figure*}

For the experiment over a month, $b_\mathrm{RH}$ is measured, the strain energy release rate is considered constant $G\sim10.8\,\mathrm{J}/\mathrm{m}^2$. To calculate $V(t)$ with the fit of the peeling master curve presented above, we need to extrapolate the rescaling factor $a_\mathrm{RH}$ from the measurements of RH. In Fig.\,S\ref{fig:SI_aRH_over_a_month}(a), the same spline interpolation (dashed line) on the hygrometry data (blue symbols) as in Fig.\,S\ref{fig:SI_water_content} is used  to extrapolate the water content $W_\mathrm{c}(t)$ corresponding to  the peeling experiments over a month (square symbols).

These extrapolated values of water content $W_\mathrm{c}(t)$ are then used to calculated with the time-humidity superposition principle (Eq.~(1)) the rescaling factor $a_\mathrm{RH}(t)$ for the peeling experiment over a month (see Fig.~S\ref{fig:SI_aRH_over_a_month}(b)). \\

\begin{figure*}[h!]
    \centering
    \includegraphics[height=8cm]{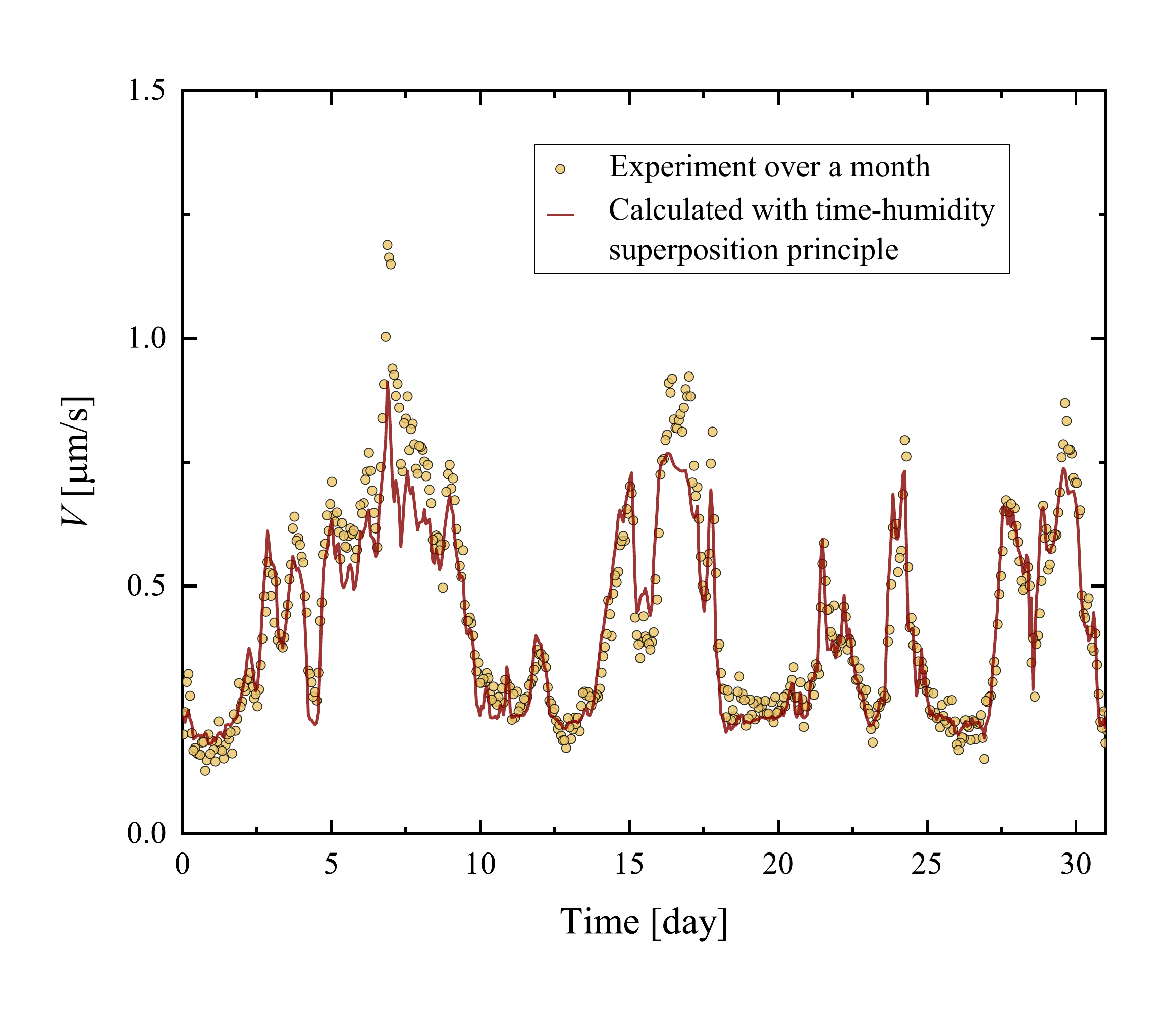}
    \caption{Unrolling of a suspended roll of tape over a period of one month. Velocity of peeling $V$ of the roll of tape as a function of time. The red line is the velocity of peeling calculated for the time-humidity superposition principle. \footnotesize}
    \label{fig:SI_compare_calculation_exp}
\end{figure*}

Last, with the fit of the peeling master curve and the time-humitidy superposition principle, the velocity of peeling $V(t)$ over a month can be calculated, as presented in Fig.~1(b) (red) and shown again in Fig.~S\ref{fig:SI_compare_calculation_exp} for sake of clarity. Calculated values are in good agreement with the experiment data, revealing the accuracy of the peeling master curve.

\clearpage
\section{Stretching of fibrils}

\begin{figure*}[h!]
    \centering
    \includegraphics[height=9cm]{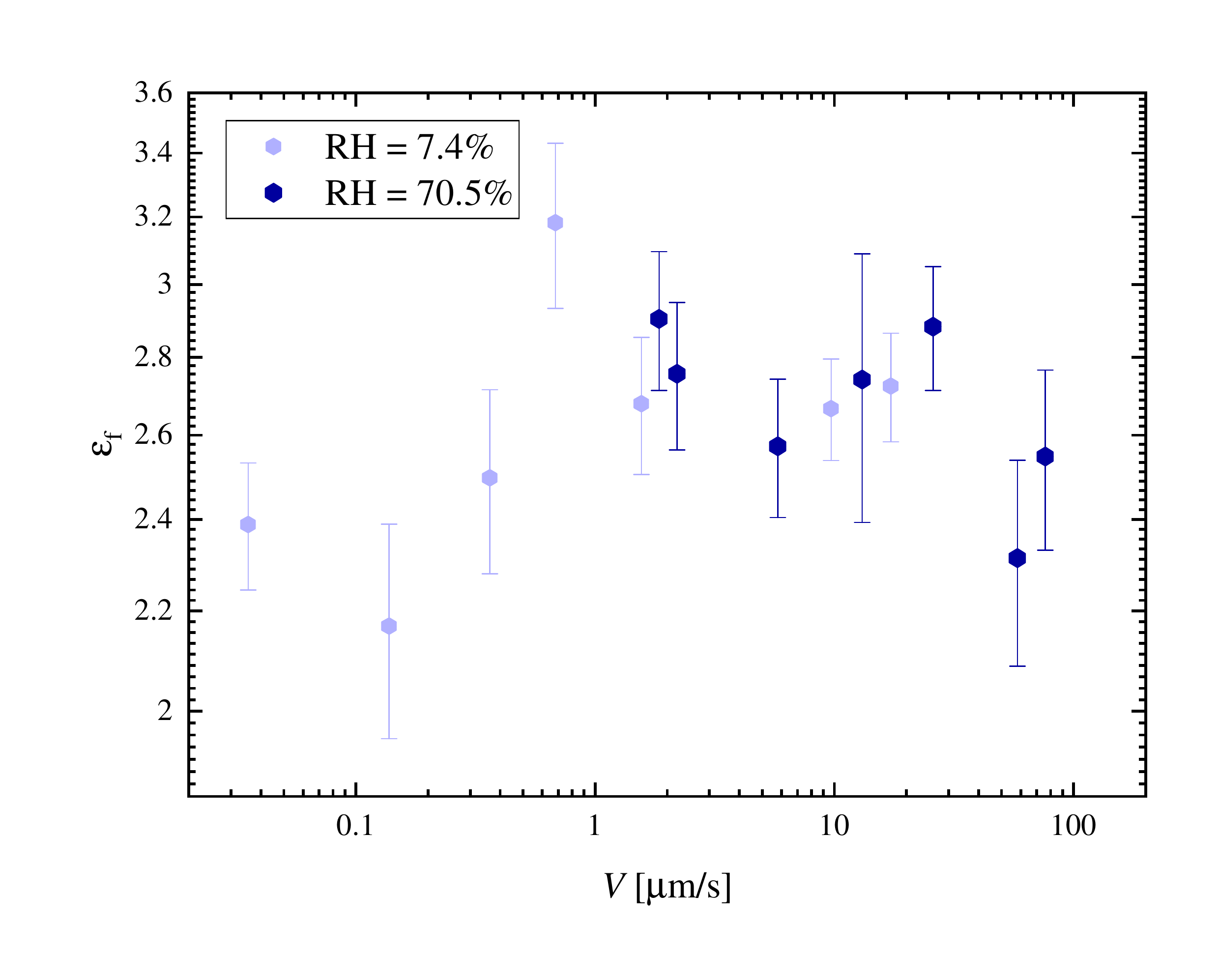}
    \caption{\textcolor{black}{Maximum fibril stretch $\epsilon_\mathrm{f}$ at debonding as a function of the peeling velocity $V$  for two different relative humidities} \footnotesize}
    \label{fig:SI_fibril stretch}
\end{figure*}

\textcolor{black}{As shown in Fig. S1(a), complex fibrils also form around the debonding region in our peeling experiments. The maximum fibril stretch $\epsilon_\mathrm{f}$ is defined as $\mathrm{ln}(a_\mathrm{f}/e)$, with $a_\mathrm{f}$ the maximal fibril length at debonding and $e=28\,\upmu$m the thickness of the adhesive. Fig.S\ref{fig:SI_fibril stretch} shows the maximum fibril stretch $\epsilon_\mathrm{f}$ at debonding as a function of the peeling velocity $V$  for two different relative humidities.  Within the limitations of our setup, the stretching of the fibrils seems to be independent of the peeling velocity $V$ and the relative humidity.}

\textcolor{black}{Chopin \textit{et al.}~\cite{chopin_nonlinear_2018, villey_rate-dependent_2015} recently showed that the adherence curve $G(V)$ presents two contributions: a rate-dependent linear viscoelastic factor and a nonlinear factor linked to the fibril stretching. They showed that a polymer for which the stretching of the fibrils is rate independent, the adherence curve is mainly due to linear viscoelastic rheology.
As the tape we studied also presents a rate independent stretching of the fibrils, the linear viscoelasticity dominates in the adherence curve. Thus, the 'time-humidity' superposition principle is applicable to rationalize the peeling experiments. Note that the same rate independence of the stretching of fibrils have been reported for the Scotch tape 600 by Villey~\cite{villey_-situ_2017}, allowing us to compare the peeling mechanism of the tape 810 to this one.}

\end{singlespace}
\bibliographystyle{apsrev4-2}
\bibliography{SI_PRL}